\documentclass[reprint,amsmath,amssymb,aps,superscriptaddress,prl]{revtex4-2}
\usepackage[dvipdfmx]{graphicx}
\usepackage[dvipdfmx]{color}
\usepackage{booktabs}
\usepackage{lineno}
\usepackage{siunitx}
\begin{document}

%\preprint{}

\title{Picosecond Trajectory of Two-dimensional Vortex Motion in FeSe$_{0.5}$Te$_{0.5}$ Visualized by Terahertz Second Harmonic Generation}

\author{Sachiko Nakamura}
\email{nakamura.sachiko@phys.kyushu-u.ac.jp}
%\thanks{}
\altaffiliation[\\Present address: ]{Department of Physics, Kyushu University, 744 Motooka, Nishi-ku, Fukuoka 819-0395, Japan }
\affiliation{Cryogenic Research Center, University of Tokyo, Bunkyo, Tokyo 113-0032, Japan}
\affiliation{JST, PRESTO, Kawaguchi, Saitama 332-0012, Japan}

\author{Haruki Matsumoto}
\affiliation{Department of Physics, University of Tokyo, Bunkyo, Tokyo 113-0033, Japan}
\author{Hiroki Ogawa}
\affiliation{Department of Basic Science, University of Tokyo, Meguro, Tokyo 153-8902, Japan}
\author{Tomoki Kobayashi}
\affiliation{Department of Basic Science, University of Tokyo, Meguro, Tokyo 153-8902, Japan}
\author{Fuyuki Nabeshima}
\affiliation{Department of Basic Science, University of Tokyo, Meguro, Tokyo 153-8902, Japan}
\author{Atsutaka Maeda}
\affiliation{Department of Basic Science, University of Tokyo, Meguro, Tokyo 153-8902, Japan}
\author{Ryo Shimano}
\email{shimano@phys.s.u-tokyo.ac.jp}
\affiliation{Cryogenic Research Center, University of Tokyo, Bunkyo, Tokyo 113-0032, Japan}
\affiliation{Department of Physics, University of Tokyo, Bunkyo, Tokyo 113-0033, Japan}


\date{\today}

\begin{abstract}
We have investigated the vortex dynamics in a thin film of an iron-based superconductor FeSe$_{0.5}$Te$_{0.5}$ by observing second-harmonic generation (SHG) in the THz frequency range. We visualized the picosecond trajectory of two-dimensional vortex motion in a pinning potential tilted by Meissner shielding current. The SHG perpendicular to the driving field is observed, corresponding to the nonreciprocal nonlinear Hall effect under the current-induced inversion symmetry breaking, whereas the linear Hall effect is negligible. The estimated vortex mass, as light as a bare electron, suggests that the vortex core moves independently from quasiparticles at such a high frequency and large velocity $\approx\SI{300}{km/s}$.

\end{abstract}

%\keywords{}

\maketitle


%The study of 
Quantum vortex in superconductors has gained continuing interest both from practical and fundamental points of view over decades. 
Intensive studies have been devoted to characterizing the behavior of the depinning current density~\cite{RevModPhys.66.1125,fluxpinning} and depinning frequency~\cite{PhysRevLett.16.734} for large current and high magnetic-field applications of superconductors. 
Recently, it has been predicted that a vortex at a surface of a topological superconductor accommodates a Majorana quasiparticles that leads to a zero-energy Majorana bound state (MBS)~\cite{PhysRevLett.100.096407, PhysRevB.61.10267}, the presence of which has been suggested in iron-based superconductors~\cite{Wang2018,Machida2019,Kong2019,Wang2020,Zhu2020,Fan2021,Kong2021,Wang2022}. 
 
Much attention has also been focused on nonreciprocal responses in inversion-broken superconductors exhibiting nonreciprocal electric transport phenomena~\cite{Wakatsuki2017,Itahashi2020,PhysRevResearch.2.023127,PhysRevMaterials.4.074003,PhysRevResearch.2.042046,Itahashi2022} and nonreciprocal critical currents or magnetic fields~\cite{Ando2020,Miyasaka2021}. 
Extensive theoretical investigations regarding the mechanism of the nonreciprocal responses have been developed~\cite{Fischer2023,PhysRevLett.121.157003,PhysRevLett.121.026601,PhysRevB.98.054510,PhysRevB.100.220501,PhysRevB.105.024308,PhysRevLett.128.037001,PhysRevB.107.024513}. 

Recently, nonreciprocal responses originated from the vortex motion were found to appear in quasi-optical, specifically terahertz (THz), frequencies in a dirty limit superconductor NbN under supercurrent injection. Here the supercurrent acts as an inversion and time-reversal symmetry breaking field, giving rise to gigantic second harmonic generation (SHG) ~\cite{PhysRevLett.125.097004}. 
At such high frequencies, the dynamics of a vortex was shown to be dominated by the motion of a single vortex core irrespective of vortex-vortex interactions.

In this Letter, we have applied this THz-SHG detection technique to investigate the vortex dynamics in an iron-based superconductor FeSe$_{1-x}$Te$_{x}$ (FST) under the diamagnetic shielding current induced through the Meissner-Ochsenfeld effect. 
Combined with the polarization- and phase-resolved measurements, we have visualized the two-dimensional trajectory of a moving vortex in a thin film of FST.

According to previous studies of microwave and DC transport measurements, a vortex in FST has properties of that in clean-limit superconductors where quasiparticles are rarely scattered in a vortex core behaving as a quantum well. 
Accordingly, the vortex mass becomes very heavy since most of normal conducting electrons accompany the vortex motion, and the long scattering time also leads to a gyroscopic force perpendicular to the vortex motion~\cite{RevModPhys.66.1125,LO1975,Kopnin1976,KopninBook}. 
The gyroscopic force has been experimentally observed through a large vortex Hall conductivity in FST~\cite{PhysRevB.91.054510,Okada2021,Ogawa2023}. 
In the present case, however, the fast-moving vortex in the FST film has negligibly small vortex Hall responses as demonstrated by the direct observation of the ultrafast vortex trajectory. By contrast, a \textit{nonreciprocal nonlinear} Hall effect is observed as a perpendicular SHG component, reflecting the supercurrent-induced inversion symmetry breaking. Furthermore, the high-frequency THz measurements reveal that the vortex mass $m_\text{v}$ in the FST film is as light as the bare electron mass, as a consequence of the feature that the mass term $(m_\text{v}\ddot{x}\propto \omega_0^2$) exceeds the viscous term ($\propto\dot{x}\propto \omega_0$) in the equation of motion of a vortex oscillating at THz frequencies $\omega_0$ where $x$ represents the displacement of the vortex. 

\begin{figure}[b]
\includegraphics[width=0.47\textwidth]{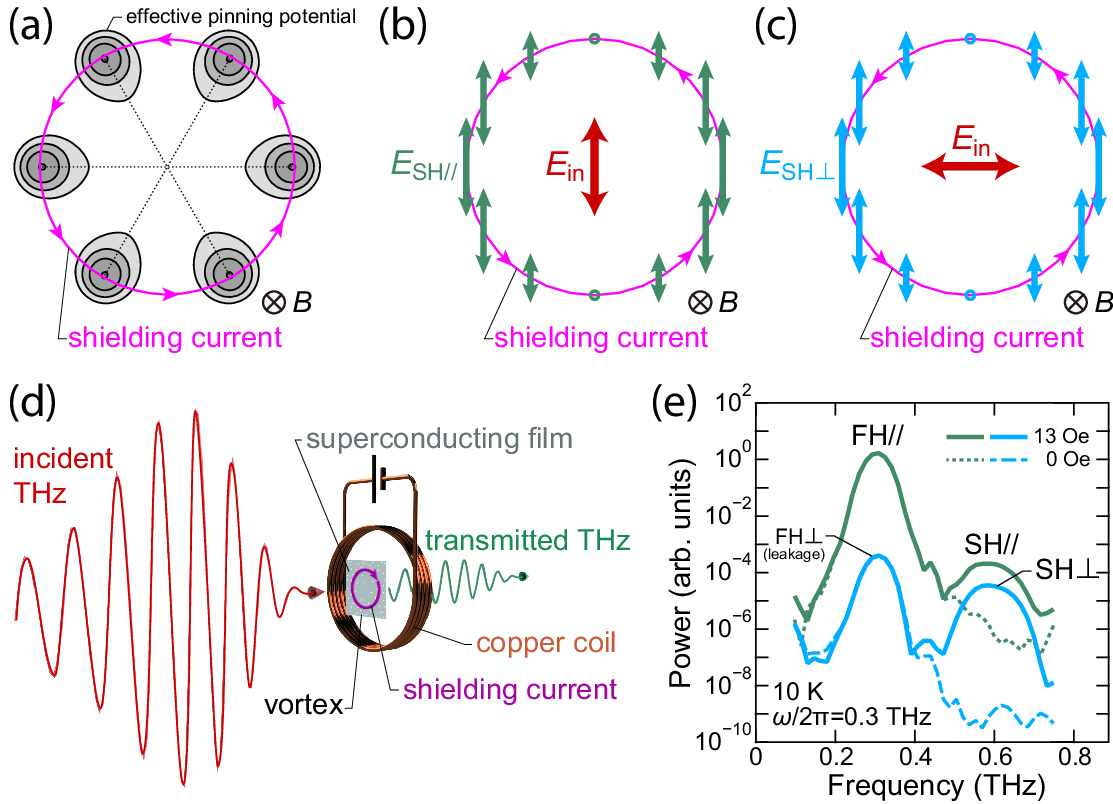}%
\caption{(a) Effective pinning potential tilted by a circular shielding current. (b,c) Position dependence of SHG whose polarization is (b) parallel ($E_{\text{SH}\parallel}$) or (c) perpendicular ($E_{\text{SH}\perp}$) to the incident polarization ($E_{\text{in}}$). 
(d) Schematic view of the THz-SHG experiments with a shielding current. 
(e) Power spectra of the transmitted THz when the incident and detected polarizations are parallel ($\parallel$) or perpendicular ($\perp$), and with (solid lines) or without (broken lines) applying a magnetic field. FH and SH stand for fundamental and second harmonics, respectively. 
\label{fig1}}
\end{figure}%\clearpage

First, we report the observation of THz-SHG caused by the vortex motion in a thin film of FeSe$_{0.5}$Te$_{0.5}$. 
In general, SHG occurs when the inversion-symmetry is broken. Therefore SHG is usually observed in materials with noncentrosymmetric structures~\cite{Tokura2018}. However, in superconductors, the inversion symmetry can also be broken by supercurrent injection~\cite{PhysRevLett.125.097004}. 
In this work, instead of the external current adopted in previous studies~\cite{PhysRevLett.122.257001,PhysRevLett.125.097004}, we propose a contact-free method for supercurrent injection that utilizes a superconducting shielding current due to the Meissner-Ochsenfeld effect induced by an external magnetic field. The local current density is expected to be as high as the critical current density~\cite{PhysRevLett.8.250,RevModPhys.36.31}. 
For thin film geometries, the magnetic field should be applied normal to the film to induce an in-plane circular supercurrent, which also generates vortices penetrating the film. 

The shielding current is expected to circulate in the film as depicted in Fig.~\ref{fig1}(a). 
The circumferential current drives vortices in the radial direction through the Lorentz force, 
which can be represented as a position-dependent effective potential whose contours are schematically drawn for 6 different positions in the figure. 
The current-induced symmetry breaking can cause two kinds of SHG: parallel SHG ($E_{\text{SH}\parallel}$) and perpendicular SHG ($E_{\text{SH}\perp}$) distinguished by the polarization of the SH with respect to that of the incident THz field ($E_{\text{in}}$)~\cite{Kleinman1962}. 
When the incident polarization is fixed, the intensity of SHG depends on the position and is largest when $E_\text{SH}$ is parallel to the current, as depicted in Fig.~\ref{fig1}(b) and (c) for parallel and perpendicular SHG, respectively. 

The SHG was measured using the THz time-domain spectroscopy (THz-TDS) techniques in a transmission geometry, as schematically illustrated in Fig.~\ref{fig1}(d), where the magnetic field is applied using a flat copper magnet (coil constant: 130\,Oe/A; inhomogeneity: $\approx6$\%). 
Power spectra of the transmitted THz pulses are shown in Fig.~\ref{fig1}(e), where the SH peak appears at 0.6\,THz in a magnetic field of 13\,Oe (solid lines) but is absent without the field (broken lines). 
The incident multicycle THz pulse was prepared from the intense monocycle pulse generated by the tilted-pulse front method with a LiNbO$_3$~\cite{Hebling2002,Watanabe2011} using bandpass filters.  
We used two narrow-band THz sources whose center frequencies $\omega/2\pi$ were 0.3 and 0.48\,THz, and their typical peak values of the electric field were 8.6 and 13.7\,kV/cm, respectively, which are below the saturation threshold of the THz-SHG.  
A regenerative amplified Ti:sapphire laser system with 800\,nm center wavelength, 100\,fs pulse duration, 1\,kHz repetition rate, and pulse energy of 4\,mJ was used as a light source. 
The THz intensity and polarization angle were controlled by wire grid polarizers and half-wave plates. 
The transmitted THz pulse after the sample was detected by electro-optic (EO) sampling using a ZnTe crystal. 
As a sample, we used an epitaxial FeSe$_{0.5}$Te$_{0.5}$ film of 38\,nm in thickness grown on a 500-\si{\micro\metre}-thick CaF$_2$ substrate by a pulsed laser deposition method. The critical temperature $T_\text{c}$ is 16.5\,K, and the critical current density $J_c$ at 10\,K is expected to be $\approx\SI{1}{MA/cm^2}$ estimated from the direct measurement for another film of similar thickness (44\,nm). 

\begin{figure}
\includegraphics[width=0.48\textwidth]{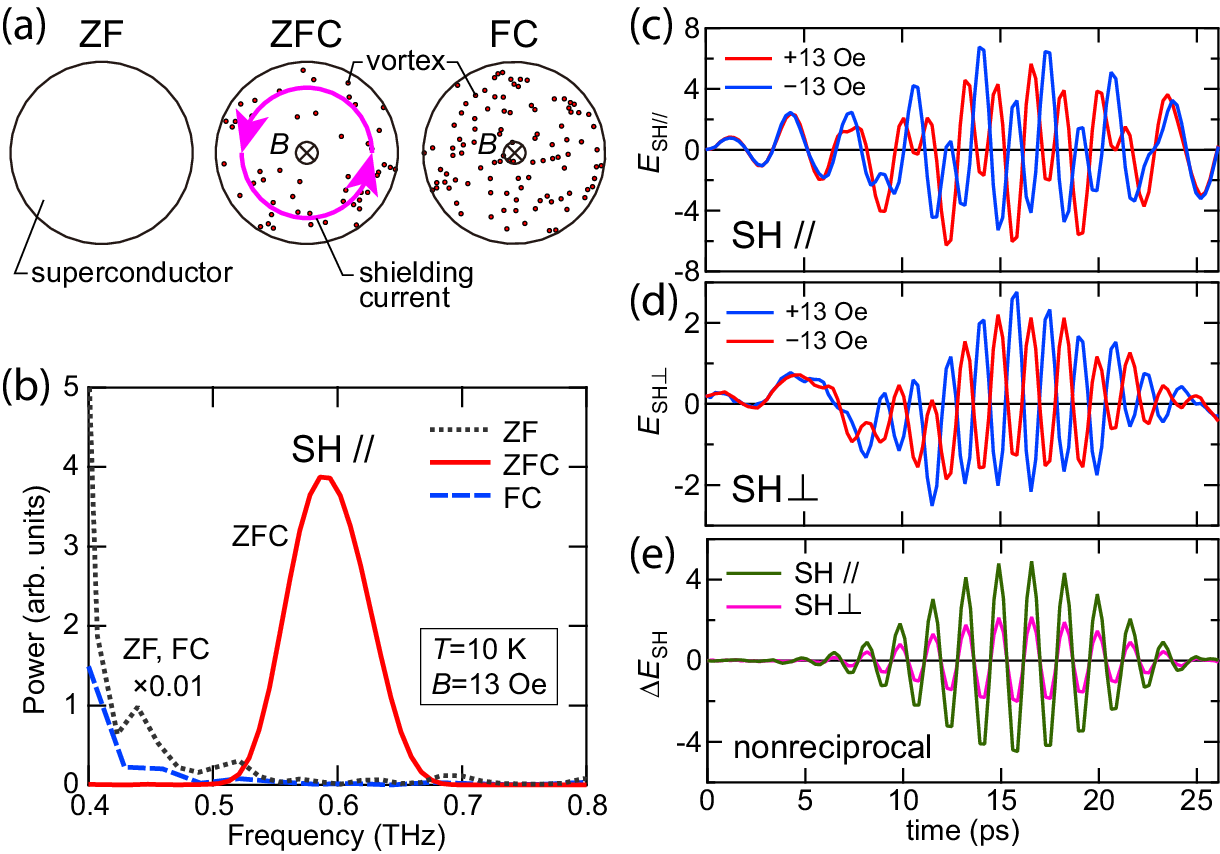}%
\caption{(a) Schematic distribution of vortices and shielding current in a superconductor for zero magnetic field(ZF), zero field cooling(ZFC), and field cooling(FC). (Details of the definition are described in the text).
(b) Power spectra of transmitted THz pulses at $T=10$\,K and $\omega/2\pi=0.3$\,THz. (c,d) Waveforms of transmitted THz pulses after ZFC at 10\,K for (c) parallel and (d) perpendicular SHG measured with detection band-pass filters whose center frequency is 0.6\,THz. (e) Nonreciprocal components $\Delta E_\text{SH} = \left[E_\text{SH}(+13\text{\,Oe})-E_\text{SH}(-13\text{\,Oe})\right]/2$ calculated from waveforms plotted in (c) and (d). \label{fig2}}
\end{figure}

Figure~\ref{fig2}(a) shows the schematic distribution of vortices and shielding current in the superconducting film for different magnetic field hysteresis. 
When a film is cooled down in the zero magnetic field (ZF: Zero Field), there are neither vortices nor a shielding current. 
If we apply a magnetic field $B$at a temperature below $T_c$ (ZFC: Zero Field Cooling), vortices enter the film from the outside and a circular shielding current appears. 
If we apply a magnetic field $B$ at a temperature above $T_c$ before cooling down (FC: Field Cooling), vortices distribute almost uniformly in the film without shielding current. 
Among these three situations, the SH peak only appears for ZFC where both vortices and shielding current exist as depicted in Fig.~\ref{fig2}(b). 
It has been demonstrated that the SHG by vortex motion observed in NbN shows nonreciprocal characteristics, namely, the SHG waveform flips the sign by reversing the direction of the current~\cite{PhysRevLett.125.097004}. 
We observed the same behavior for both parallel and perpendicular SHG when we reverse the shielding current by applying a magnetic field of the opposite sign as shown in Figs.~\ref{fig2}(c--e) where Fig.~\ref{fig2}(e) exhibits nonreciprocal components consisting only of SH components. 
%Flipping the magnetic field changes the sign of vortices and the direction of the vortex motion, but it does not affect the electric response. 

\begin{figure}
\includegraphics[width=0.48\textwidth]{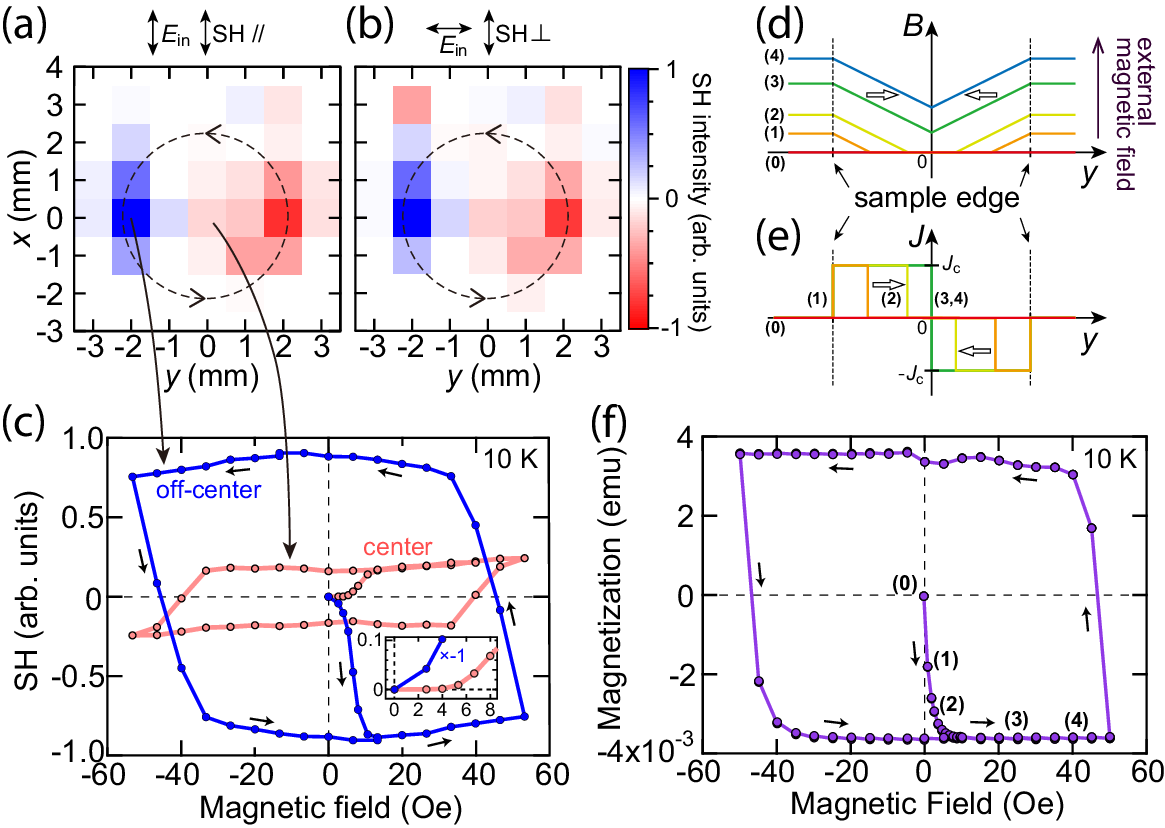}%
\caption{(a,b) Color map of (a) parallel and (b) perpendicular SHG at $T=10$\,K and $\omega/2\pi=0.48$\,THz where the detection polarization is along the $x$-axis. 
The color scale is shown on the right. 
(c) Hysteresis loops of the parallel SHG for two different spots in the sample. The inset is a closeup around the origin.
(d,e) Schematic diagrams of the magnetic field $B$ and electric current $J$ in a superconducting sample when we increase the external magnetic field from zero where $J_c$ is the critical current density. (0)--(4) corresponds to the same characters in (f). 
(f) Magnetization hysteresis loop for another film of FeSe$_{0.5}$Te$_{0.5}$ of similar shape (5\,mm square and 55\,nm thick) for magnetic fields normal to the film.\label{fig3}}
\end{figure}

The position dependence of the SH signal was measured by moving the sample on the focal plane of the incident 0.48\,THz-FH beam whose spot size is $\approx1$\,mm. 
The shape of the sample is 7\,mm square secured on a metallic tapered hole with a diameter of 6\,mm. 
Both parallel and perpendicular SHG have clear and similar position dependencies as depicted in Fig.~\ref{fig3}(a) and (b), which agrees with the expected behavior shown in Figs.~\ref{fig1}(b) and (c). 
The intensity of SHG is strong when its polarization is parallel to the shielding current with the sign, indicated by red and blue, reflecting the direction of the current. For example, both SHG and current at $(x,y)=(0,-2)$ has the opposite sign of them at $(x,y)=(0,2)$ as discerned in Figs.~\ref{fig1}(a) and (b). 

The parallel SHG shows a magnetic hysteresis loop as depicted in Fig.~\ref{fig3}(c). 
The larger loop was measured at $(x,y)=(0,-2)$ and the smaller one was at the center $(0,0)$. 
With increasing the magnetic field from zero, SHG first increases in intensity and then saturates at $\approx10$\,Oe.  
When we decrease the magnetic field from 50\,Oe, the SHG immediately goes beyond zero to the saturated value with the opposite sign. 
This behavior is similar to the hysteresis loop of magnetization measured for a similar film of FeSe$_{0.5}$Te$_{0.5}$ depicted in Fig.~\ref{fig3}(f). 
Note that the largest magnetic field applied in this measurement, 50\,Oe, is far less than the huge upper critical field, $H_{c2}>$10\,T$=10^5$\,Oe of this kind of films~\cite{Tsukada2011}. 

The observed magnetization hysteresis loop is accounted for by Bean's critical state model~\cite{RevModPhys.36.31}. 
When we increase the external magnetic field from zero, as indicated by (0)--(2) in Fig.~\ref{fig3}(d--f), 
the magnetic field enters the sample from the edge with a constant field gradient as depicted in Fig.~\ref{fig3}(d). 
The circular shielding current simultaneously enters from the sample edge with a constant current density $J=\pm J_c$ as depicted in Fig.~\ref{fig3}(e). 
In response to the $B$ and $J$, the SHG measured at the edge smoothly increases the intensity while the SHG measured at the center has a zero plateau as shown in Fig.~\ref{fig3}(c) inset because neither current nor magnetic field reaches the center. 
At $\approx 5$\,Oe, the SHG at the center starts to increase because the $B$ and $J$ reach the center, and at the same time, the magnetization saturates. 
The SHG saturates at $\approx 10$\,Oe, which presumably indicates that the vortex density reaches the closest packing density for Pearl vortex~\cite{Pearl1964} which makes it difficult to use the model of an isolated vortex for THz-SHG. 

\begin{figure}
\includegraphics[width=0.48\textwidth]{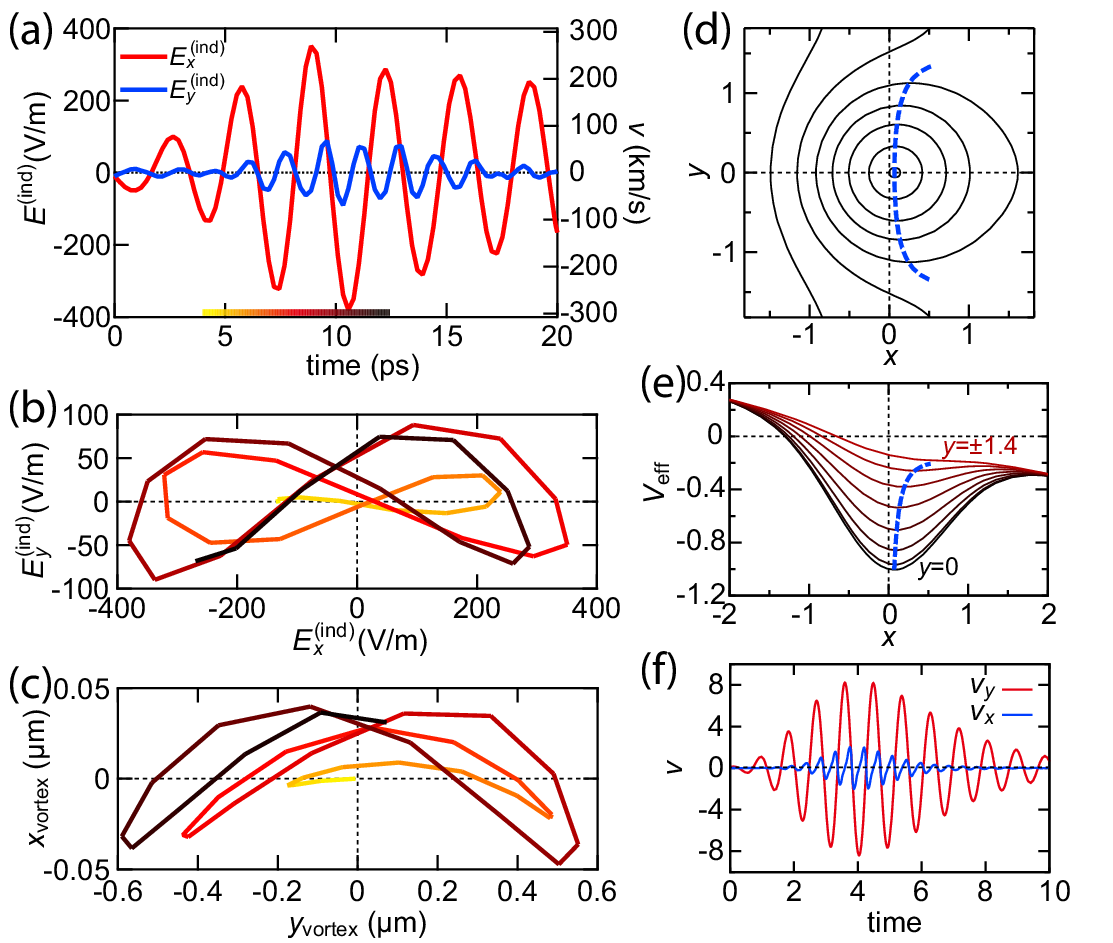}%
\caption{(a) Waveforms of the change of the transmitted THz induced by applying a magnetic field of 13\,Oe at $T=10$\,K and $\omega/2\pi=0.3$\,THz, as denoted by $E^{\text{(ind)}}$. The peak value of $E_\text{in}$ was 4.5\,kV/cm. (b) Lissajous curve for $E^{\text{(ind)}}$. The color scale is indicated on the bottom axis of (a). (c) Real-space trajectory of a vortex reconstructed from $E^{\text{(ind)}}$. (d) Contours for an effective pinning potential $V_\text{eff}$ (defined in the text) with a current-induced tilt in $x$-axis. The dashed line indicates the $x$ values of potential minima for constant $y$. (e) Solid lines indicate $V_\text{eff}$ for $y=0, \pm0.2, \cdots, \pm1.4$ and the dashed line shows their minima. (f) Velocity of a vortex calculated for the $V_\text{eff}$ when an oscillating pulse of external force is applied along $y$-axis.  \label{fig4}}
\end{figure}

As parallel SHG is explained by one-dimensional vortex motion~\cite{PhysRevLett.125.097004}, here, in order to consider the microscopic mechanism of perpendicular SHG, we visualize two-dimensional vortex motion estimated from the THz emission $E^\text{(ind)}$ as depicted in Fig.~\ref{fig4}(a). 
In practice, $E^\text{(ind)}$ is calculated as the change of the transmitted THz waveform between ZFC and ZF measured at $(x,y)=(0,-2)$. Note that the magnetic field and the shielding current are much smaller than the upper critical field and the depairing current, respectively, and thus their effects on the THz transmittance are negligible.  
The Lissajous curve of $E^\text{(ind)}$ is shown in Fig.~\ref{fig4}(b). 
The vortex velocity $\vec{v}$ can be calculated from $E^\text{(ind)}$ using the following equation: 
\begin{equation}
\vec{E}^{\text{(ind)}} =n_0 \Phi_0 \vec{z}\times \vec{v}
\end{equation}
where $n_0$ is the density of vortices, $\vec{z}$ is the unit vector normal to the film, and $\Phi_0\equiv h/(2e)$ is the magnetic flux quantum. 
The maximum velocity is $\approx\SI{300}{km/s}$ as indicated in Fig.~\ref{fig4}(a) with the right axis. 
Using the $\vec{v}$, we reconstruct the vortex trajectory as depicted in Fig.~\ref{fig4}(c), where the vortex mainly oscillates in $y$-axis and the overall trajectory exhibits a parabola-like curve. 

The trajectory first reveals that the vortex motion is nearly perpendicular to the incident THz field $E_\text{in}$ along $x$-axis, where the motion amplitude in $x$-axis at the FH frequency is only 5\% of that in $y$-axis. 
In other words, the vortex moves parallel to the Lorentz force due to the supercurrent induced by $E_\text{in}$. 
If the quasiparticles in the core dominate the vortex motion, the vortex would move almost perpendicular to the driving force~\cite{PhysRevB.44.9667,Ogawa2023}. 
The large oscillation amplitude ($\approx \SI{1}{\micro m}$) indicates the small vortex viscosity $\eta$ as estimated to be $\approx \SI{1e-11}{kg/m\cdot sec}$~\cite{sm}, which is by far smaller than that expected for clean vortex cores with a plenty of quasiparticles ($\approx \SI{1e-7}{kg/m\cdot sec}$~\cite{Okada2021}). 
Therefore, the observed trajectory indicates that quasiparticles trapped by the core in the stationary state have little effect on the high speed motion of the vortex. 

To explain the parabola-like trajectory, we consider a moving vortex under a two-dimensional pinning potential $V_\text{pin}$ of Gaussian shape. 
Under an electric current along the $y$-axis, the effective potential $V_\text{eff}$ is tilted in $x$-axis and has a function form of 
\begin{equation}
V_\text{eff}\equiv V_\text{pin}-ax=-\exp\left[-(x^2+y^2)\right]-ax
\end{equation}
whose contours are depicted in Fig.~\ref{fig4}(d) for $a=0.14$. 
The vortex feels the oscillating driving force along $y$-axis.  
Here we plot $V_\text{eff}$ as a function of $x$ for different values of $y$ in Fig.~\ref{fig4}(e), where the potential minimum moves along the dashed curve, i.e., the oscillation in $y$-axis can induce the oscillation in $x$-axis. 
The potential minimum forms a parabola-like curve on the $x$-$y$ plane depicted as the thick dashed line in Fig.~\ref{fig4}(d), which has similar characteristics to the observed vortex trajectory shown in Fig.~\ref{fig4}(c). 
In the lowest-order approximation, $x$ is proportional to $y^2$ on the curve, and thus the oscillation frequency along the $x$-axis is twice the frequency along the $y$-axis. 
We depict a typical waveform of vortex velocity in Fig.~\ref{fig4}(f) assuming that the vortex oscillates along the curve of the potential minimum, where $v_y$ and $v_x$ corresponds to $E_x^\text{(ind)}$ and $E_y^\text{(ind)}$ in Fig.~\ref{fig4}(a). 
It is worth noting that the vortex moves along $x$ direction (Fig.~\ref{fig4}(c)) in a slow time scale, which can be viewed as a rectification effect of the vortex motion driven by the ac driving field.

Here we discuss the vortex mass $m_\text{v}$. 
The mass of a clean vortex core has been considered to be of the order of the total mass of electrons in the core~\cite{PhysRevLett.81.3952} ($\approx 10^4 m_\text{e}$~\cite{sm}), since the quasiparticles bound in the core are considered to move with the core~\cite{Kopnin1978,Volovik1997}. 
However, based on the present THz-SHG results, we estimate $m_\text{v}$ to be $\approx m_\text{e}$~\cite{sm} which is as small as that in the dirty-limit NbN~\cite{PhysRevLett.125.097004}.  
Considering the results of the mass and viscosity, it is likely suggested that the quasiparticles originally trapped in the stationary vortex core cannot follow the fast movement of the vortex driven by the high-intensity THz pulse.  
Since the FH transmittance does not increase over the duration of the THz pulse, it appears that the quasiparticles liberated from the vortex core do not destabilize the superconductivity. 
To reconcile this, recently discovered phenomenon where fast-moving vortices flowing at $\approx\SI{100}{km/s}$ form a river~\cite{PhysRevB.105.214507} is suggestive. 
It has indicated that liberated quasiparticles are immediately retrapped by the next vortex flowing from upstream in tens of picoseconds and the entire superconductivity is not destroyed~\cite{PhysRevB.102.024506,PhysRevApplied.17.034072,PhysRevB.103.134511}. 
In our case, the vortex core oscillates only during the duration of the incident THz pulse, typically 20--$\SI{30}{ps}$, therefore, once the THz pulse has passed, the core that ceased its oscillation can retrap the quasiparticles before spreading. 

Finally, we briefly address the Majorana bound state (MBS) which has been reported to exist in a vortex core at low temperatures below 3\,K in FST~\cite{Wang2018,Machida2019,Kong2019,Wang2020,Zhu2020,Fan2021,Kong2021,Wang2022}. 
Unlike charged quasiparticles accelerated by electric fields~\cite{LO1975}, it is a nontrivial problem as to whether the charge-less MBS are deconfined from the vortex core oscillating in an ultrafast ($\sim\si{ps}$) time scale and what the energization mechanisms are~\cite{PhysRevB.107.205420}. Thus it is tempting to applying the demonstrated scheme of THz-SHG at lower temperatures to study the dynamics of a vortex accommodated with the MBS which we remain as a future work. 

In summary, we have demonstrated time-resolved visualization of the trajectory of a fast-moving vortex in a thin film of iron-based superconductor FeSe$_{0.5}$Te$_{0.5}$ using phase-resolved THz-TDS techniques. 
The polarization-resolved measurements revealed the THz-SHG with polarization perpendicular to that of the incident THz pulse, and its origin is attributed to the nonreciprocal nonlinear Hall effect as unveiled from the vortex trajectory, reflecting the supercurrent-induced inversion symmetry breaking. 
The properties of the fast-moving vortex are estimated from the analysis of THz-SHG, providing a small mass and viscosity of the vortex. 
This result suggests that most quasiparticles originally bound in the static vortex core do not catch up with the fast movement of the vortex. 
We have also demonstrated new schemes to visualize shielding supercurrents through the observation of phase-resolved THz-SHG, which may encourage spectroscopic studies with supercurrent injection on various superconductors. 

\begin{acknowledgments}
This work was supported in part by JSPS KAKENHI (Grants No. 20K14408), by JST PRESTO Grant No. JPMJPR2108, Japan, and by JST CREST Grant No. JPMJCR19T3, Japan.
\end{acknowledgments}

%\bibliography{shieldingcurrent_bib}
%
\end{document}